\RequirePackage[T1]{fontenc}
\documentclass[12pt]{article}
\pdfoutput=1
\usepackage[height=8.85in,width=6.45in]{geometry}

\usepackage{times}

\usepackage[utf8]{inputenc}
\usepackage{amsmath}
\usepackage{amssymb}
\usepackage{mathtools}
\numberwithin{equation}{section}
\usepackage{slashed}
\usepackage{braket}
\usepackage[svgnames,psnames]{xcolor}
\usepackage[colorlinks,citecolor=DarkGreen,linkcolor=FireBrick,linktocpage]{hyperref}
\usepackage{cite}
\usepackage{graphicx}
\usepackage{tikz}
\usepackage{tikz-cd}

\usepackage{courier}
\usepackage{bm}
\usepackage{subfig}
\usepackage{ascmac}
\usepackage{dashbox}

\usepackage{mdframed}

\renewenvironment{figure}[1][]{
  \begin{originalfigure}[#1]
    \begin{mdframed}[linecolor=black!0,backgroundcolor=black!1]
}{
    \end{mdframed}
  \end{originalfigure}
}
\renewenvironment{table}[1][]{
  \begin{originaltable}[#1]
    \begin{mdframed}[linecolor=black!0,backgroundcolor=black!1]
}{
    \end{mdframed}
  \end{originaltable}
}

\usepackage{colortbl}
\definecolor{lightyellow}{rgb}{1.0, 0.95, 0.7}
\definecolor{lightblue}{rgb}{0.7, 0.9, 1.0}
\definecolor{lightpink}{rgb}{1.0, 0.85, 0.95}
\definecolor{lightgreen}{rgb}{0.7, 1.0, 0.4}
\definecolor{blue}{rgb}{0.0, 0.4, 1.0}
\definecolor{green}{rgb}{0.2, 0.7, 0.2}
\definecolor{Blue}{rgb}{0,0,1}
\definecolor{darkgreen}{rgb}{0.,0.6,0.}

\newcommand*{\bZ}{\mathbb{Z}}
\newcommand*{\bR}{\mathbb{R}}

\newcommand*{\cH}{\mathcal{H}}

\def\Nequals#1{$\mathcal{N}{=}#1$}

\def\Hom{\mathop{\mathrm{Hom}}\nolimits}

\def\Inv{\mathop{\mathrm{Inv}}\nolimits}

\def\str{\mathsf{S}}
\usepackage{arydshln}

\begin{document}

\begin{titlepage}

\begin{flushright}
IPMU-20-0101
\end{flushright}

\vskip 3cm

\begin{center}

{\Large \bfseries $SL(2,\bZ)$ action on QFTs with $\bZ_2$ symmetry\\[1em]
and the Brown-Kervaire invariants}

\vskip 1cm
Lakshya Bhardwaj$^{1,2}$,
Yasunori Lee$^3$, 
and Yuji Tachikawa$^3$
\vskip 1cm

\begin{tabular}{ll}
$^1$ & Mathematical Institute, University of Oxford,\\
& Andrew Wiles Building, Woodstock Road, Oxford, OX2 6GG, UK\\
$^2$ & Department of Physics, Harvard University, \\
& 17 Oxford St, Cambridge, MA 02138, USA\\
$^3$ & Kavli Institute for the Physics and Mathematics of the Universe (WPI), \\
& University of Tokyo,  Kashiwa, Chiba 277-8583, Japan
\end{tabular}

\vskip 1cm

\end{center}

\noindent 
We consider an analogue of Witten's $SL(2,\bZ)$ action on three-dimensional QFTs with $U(1)$ symmetry
for $2k$-dimensional QFTs with $\bZ_2$ $(k-1)$-form symmetry.
We show that the $SL(2,\bZ)$ action only closes up to a multiplication by an invertible topological phase
whose partition function is the Brown-Kervaire invariant of the spacetime manifold.
We interpret it as part of the $SL(2,\bZ)$ anomaly of the bulk $(2k+1)$-dimensional $\bZ_2$ gauge theory.

\end{titlepage}

\setcounter{tocdepth}{2}
\tableofcontents

\bigskip

\section{Introduction and summary}
In \cite{Witten:2003ya} Witten introduced the $SL(2,\bZ)$ action on 3d quantum field theories (QFTs) with $U(1)$ symmetry given as follows.
Let $Q$ be a 3d theory with $U(1)$ symmetry whose partition function is $Z_Q[A]$, where $A$ is the background $U(1)$ gauge field.
Then the $S$ and $T$ operations are defined as \begin{align}
Z_{SQ}[A] &= \int [Da] \exp\left(2\pi i  \int_{M_3} \frac{a}{2\pi} d \frac{A}{2\pi}\right) Z_Q[a],\label{U1S}\\
Z_{TQ}[A] &= \exp\left(2\pi i \cdot \frac12 \int_{M_3} \frac{A}{2\pi} d \frac{A}{2\pi}\right) Z_Q[A].\label{U1T}
\end{align}
The prefactor in the definition of $TQ$ is a level-1 $U(1)$ Chern-Simons term\footnote{%
In some literature, including \cite{Witten:2003ya}, a different normalization was used so that this corresponds to level-1/2.
} and therefore requires that the spacetime manifold $M_3$ is endowed with a spin structure \cite{Witten:2003ya,Belov:2005ze}.

A careful argument given in \cite{Witten:2003ya} shows that one has \begin{align}
Z_{S^2Q}[A]&=Z_Q[-A],\\
Z_{(ST)^3Q}[A]&=\left(
\int [Da]
\exp\left(2\pi i \cdot \frac12 \int_{M_3} \frac{a}{2\pi} d \frac{a}{2\pi}\right)
\right) Z_Q[A], \label{U1ST3}
\end{align} 
where the factor in the parenthesis of \eqref{U1ST3} is the level-1 $U(1)$ Chern-Simons theory.
These equations imply that the operations $S$ and $T$ satisfy the relations \begin{equation}
S^2 = C, \qquad (ST)^3= Y
\label{U1anom}
\end{equation}
where $C$ is the charge conjugation of $U(1)$ symmetry and 
$Y$ is (the multiplication or equivalently the stacking by) the level-1 $U(1)$ Chern-Simons theory.
Therefore the operations $S$ and $T$ form the group $SL(2,\bZ)$, up to a multiplication by $Y$.

In \cite{Witten:2003ya} it was  found that the level-1 $U(1)$ Chern-Simons theory $Y$ has a one-dimensional Hilbert space on any closed spatial manifold. 
Such QFTs are now known as invertible phases following \cite{Freed:2004yc}.
Invertible phases form an Abelian group under multiplication,
and 3d spin invertible phases are known to form the group $\bZ$, such that the invertible theory labeled by $n\in \bZ$ has a boundary mode with the chiral central charge $c=n/2$, and has the quantized thermal Hall conductivity $\kappa_{xy}=n \pi k_B^2 T/(12\hbar)$, see e.g.~\cite{Read:1999fn} and \cite[Appendix]{Kapustin:2014dxa}.
In this normalization, the level-1 $U(1)$ Chern-Simons theory is labeled by $2\in \bZ$.
Denoting the invertible theory labeled by $1$ by $X$, we have $Y=X^2$.

This makes it clear that we cannot redefine $S$ and $T$ to make the action genuinely $SL(2,\bZ)$.
Indeed, let us try redefining $\tilde S:=X^aS$ and $\tilde T:=X^bS$, and demand $\tilde S^4=1$ and $(\tilde S\tilde T)^3=1$.
This requires $4a=0$ and $3(a+b)=2$, which do not admit any solutions.

The finding of \cite{Witten:2003ya} and the progress in the understanding of invertible phases since then can be summarized thus: 
the $SL(2,\bZ)$ action on 3d spin QFTs with $U(1)$ symmetry forms a projective representation, 
where the projective phase is given by an invertible topological phase, which is the level-1 Chern-Simons theory.

In this short note, we provide an analogue of this $SL(2,\bZ)$ operation for $2k$-dimensional QFTs with $\bZ_2$ $(k-1)$-form symmetry.
For 4-dimensional  QFTs with $\bZ_2$ 1-form symmetry,
this $SL(2,\bZ)$ action was introduced in \cite[Sec.~6]{Gaiotto:2014kfa} in the spin case,
and was later extended to the non-spin case in \cite{Ang:2019txy}.
If we denote a 4d gauge theory with $Spin$ gauge group by $Q$ which has electric $\bZ_2$ 1-form symmetry, 
the $SO_+$ gauge theory corresponds to $SQ$ and the $SO_-$ gauge theory corresponds to $STQ$.
In the 4d spin case, this action is genuine and not projective.
We will consider its generalization to more general spacetime dimensions and to more general spacetime structures, 
and show that this action is projective up to a multiplication by an invertible topological phase,
whose partition function is a Brown-Kervaire invariant \cite{Brown,KlausThesis}.
This will be done in Sec.~\ref{sec:discreteSL2Z}.

In \cite{Witten:2003ya}, it was also explained that the $SL(2,\bZ)$ action on 3d QFTs with $U(1)$ symmetry comes from considering these theories on the boundary\footnote{%
The argument was presented in the context of the AdS/CFT correspondence, but it equally applies to the bulk-boundary systems.
} of the bulk 4d Maxwell theory and performing the $SL(2,\bZ)$ duality of the bulk as first described in \cite{Witten:1995gf}.
We will describe how the projectivity of the $SL(2,\bZ)$ action on 3d QFTs arises from the mixed-gravitational anomaly of the $SL(2,\bZ)$ action of the bulk as found in \cite{Seiberg:2018ntt}.
Applying the same analysis to the $SL(2,\bZ)$ action on $2k$-dimensional QFTs with $\bZ_2$ $(k-1)$-form symmetry,
we  infer the mixed-gravitational anomaly of the $SL(2,\bZ)$ action of the bulk $\bZ_2$ gauge theory.
This is the content of Sec.~\ref{sec:interpretation}.
We are not going to find the full anomaly of the $SL(2,\bZ)$ action of the bulk $\bZ_2$ gauge theory in this paper.
Determining it will be left to the future.

\section{$SL(2,\bZ)$ action and the Brown-Kervaire invariants} 
\label{sec:discreteSL2Z}
\subsection{The operations $S$ and $T$ and their compositions}
Let us consider $2k$-dimensional QFTs with $\bZ_2$ $(k-1)$-form symmetry.
We assume this $\bZ_2$ $(k-1)$-form symmetry to be anomaly-free throughout this paper.
We are going to define two operations, $S$ and $T$, acting on this class of QFTs, forming the group $SL(2,\bZ)$ which is realized projectively.

Let us denote the partition function of a theory $Q$ as $Z_Q[B]$.
Here, $B$ is the background field for $\bZ_2$ $(k-1)$-form symmetry, 
which is a degree-$k$ cohomology class $B\in H^k(M_{2k})$,
where $M_{2k}$ is the spacetime manifold.
The cohomology groups appearing below are all $\bZ_2$-valued.

For 4d theories on spin manifolds with $\bZ_2$ 1-form symmetry, these two operations were introduced in \cite[Sec.~6]{Gaiotto:2014kfa}.
They act for example on $\mathfrak{so}$ gauge theories with matter fields in the vector representation. 
Denoting the $Spin$ theory by $Q$, the $SO_+$ theory will be $SQ$ and the $SO_-$ theory will be $STQ$.
In this case there is no projectivity, as we will see later.

The definition of the $S$ operation is straightforward and is given by gauging the $\bZ_2$ symmetry: \begin{equation}
Z_{SQ}[B] = \frac{1}{|H^k(M_{2k})|^{1/2}}\sum_{b}  (-1)^{ \langle B,b\rangle } Z_Q[b].\label{S}
\end{equation}
Here, $\langle B,b\rangle := \int_{M_{2k}} B \cup b$ is the intersection pairing
and  we have introduced a  standard\footnote{%
To see that it is the correct normalization to use, consider the case when $Q$ is a completely trivial theory. 
Then $SQ$ is the $\bZ_2$ gauge theory, whose Hilbert space on $M_{2k-1}$ should have dimension $|H^k(M_{2k-1})|$.
We now evaluate the expression \eqref{S}  for $M_{2k}=S^1\times M_{2k-1}$.
Since $H^k(S^1\times M_{2k-1}) = H^k(M_{2k-1})\oplus H^{k-1}(M_{2k-1})$
and the two direct summands are Poincar\'e dual to each other, we see that the expression \eqref{S} indeed equals $|H^k(M_{2k-1})|$.
}
prefactor $|H^k(M_{2k})|^{1/2}$.
This is merely a discrete Fourier transformation on $H^k(M_{2k})$, and
one can easily check that \begin{equation}
Z_{S^2Q}[B] = Z_Q[B],
\end{equation}
as $B$ is $\bZ_2$-valued and there is no distinction between $\pm B$. Therefore we simply have $S^2=1$.

To define the $T$ operation requires more care. 
Namely, we need a quadratic refinement $q$ of the intersection pairing $\langle -,-\rangle$.
It is a $\bZ_4$-valued function which satisfies the condition \begin{equation}\label{q}
q(B+B')=q(B)+q(B')+2\langle B,B'\rangle.
\end{equation}
Note that $q(0)=0$ follows from this relation by setting $B'=0$, which then implies that \begin{equation}
q(B)=\langle B,B\rangle \pmod 2\label{qBBB}
\end{equation} by setting $B=B'$.

There are cases when a $\bZ_2$-valued quadratic refinement $\bar q(B)$ satisfying \begin{equation}
\bar q(B+B')=\bar q(B)+ \bar q(B')+\langle B,B'\rangle
\end{equation}
is available,
but it is a special case of a $\bZ_4$-valued quadratic refinement where $q(B)=2\bar q(B)$.

Quadratic refinements do not come for free. 
We pick a spacetime structure (such as an orientation, spin, or pin$^\pm$, appropriate for the chosen dimension $d=2k$ as we detail below) such that a quadratic refinement is available,
and we restrict our attention to QFTs which include that particular spacetime structure as part of the data.
We denote the chosen structure by $\str$ when necessary.

We then define the operation $T$ as follows: \begin{equation}
Z_{TQ}[B] = i^{q(B)} Z_Q[B].
\end{equation}
It is an order four operation in general; $T^4=1$.
When $q(B)$ is always even and comes from a $\bZ_2$-valued quadratic refinement, 
it is an order two operation.

Let us now compute $STS$: \begin{align}
Z_{STSQ}[B] &=\frac1{|H^k(M_{2k})|} \sum_{b,b'} (-1)^{\langle B,b\rangle} i^{q(b)} (-1)^{\langle b,b'\rangle} Z_Q[b']\notag \\
&=\frac1{|H^k(M_{2k})|} \sum_{b,b'} i^{q(b+B+b')} i^{-q(B+b')} Z_Q[b']\notag\\
&=\left(\frac1{|H^k(M_{2k})|^{1/2}} \sum_{\tilde b} i^{q(\tilde b)} \right)
\left(
i^{-q(B)} \frac1{|H^k(M_{2k})|^{1/2}} \sum_{b'} (-1)^{\langle B,b'\rangle} i^{-q(b')} Z_Q[b']
\right)\notag\\
&= Z_Y Z_{T^{-1}ST^{-1}Q} [B],
\end{align}
where \begin{equation}
Z_Y:=\frac1{|H^k(M_{2k})|^{1/2}} \sum_{\tilde b} i^{q(\tilde b)}.
\end{equation}
We therefore see that we have \begin{equation}
(ST)^3=Y,
\end{equation}
where $Y$ is (the multiplication by) the $\bZ_2$ gauge theory whose action is $q(b)$ with the partition function given by $Z_Y$.

\subsection{Brown-Kervaire invertible field theories}
The sum $Z_Y$ is known to be $\str$-bordism invariant \cite{Brown}
and is called the Brown-Kervaire invariant of the quadratic refinement $q$ of the intersection form $\langle -,-\rangle$.
This is in general an eighth root of unity, so it is $\bZ_8$-valued \cite{Brown}.
Furthermore, $Z_Y$ becomes $\pm1$ when $q$ is even and comes from a $\bZ_2$-valued quadratic refinement $\bar q$.
In this case, $Z_Y$ is the Arf invariant of $\bar q$.

In either case, the partition function $Z_Y$ is always invertible.
Therefore this theory $Y$ is an invertible topological field theory.
The  bordism invariance of $Z_Y$, mathematically proved in \cite{Brown} long time ago, agrees with the physics expectation that invertible topological phases are bordism invariant \cite{Kapustin:2014dxa}, which was recently proved in \cite{Yonekura:2018ufj} in general.

We can also directly check that the dimension of the Hilbert space is one. 
Consider the case when $M_{2k}=M_{2k-1}\times S^1$. 
Decomposing $b\in H^k(M_{2k})$ as 
\begin{equation}
b=u\oplus v \in H^{k}(M_{2k-1})\oplus H^{k-1}(M_{2k-1}),
\end{equation}
any quadratic refinement is of the form \begin{equation}
q(b)=2\int_{M_{2k-1}} (u\cup v  + u\cup c + c'\cup v)
\end{equation}
where $c\in H^{k-1}(M_{2k-1})$ and $c'\in H^{k}(M_{2k-1})$ are elements dependent on $q$.
Before imposing the gauge invariance, the basis of the Hilbert space on $M_{2k-1}$ is labeled by $u\in H^k(M_{2k-1})$, and \begin{equation}
i^{q(b)}=(-1)^{ \int_{M_{2k-1}} (u\cup v  + u\cup c + c'\cup v) }
\end{equation} tells us how a gauge transformation by $v\in H^{k-1}(M_{2k-1})$ acts on each state.
Only the state labeled by $c'\in H^k(M_{2k-1})$ is gauge-invariant, and we find \begin{equation}
Z_Y=(-1)^{\int_{M_{2k-1}} c'\cup c}.
\end{equation}

\subsection{Comments}
\label{sec:comments}

So far, we defined two operations $S$ and $T$, introduced the invertible phase $Y$, and showed that they satisfy the relations \begin{equation}
S^2=1,\qquad T^4=1, \qquad (ST)^3=Y.\label{mainrelations}
\end{equation}
Whether the projectivity due to $Y$ can be removed by redefinitions of $S$ and $T$ 
depends both on the properties of the theory $Y$,
and on exactly which group action we would like to consider. 

Since we consider $\bZ_2$ symmetry for which there is no distinction of $\pm B\in H^k(M_{2k})$,
it is tempting to demand that we have a projective action of $SL(2,\bZ_2)$.
In this group, however, the operation $T$ has order two, while our operation $T$ has in general order four, such that \begin{equation}
Z_{T^2Q}[B]=(-1)^{q(B)} Z_Q[B]= (-1)^{\langle B,B\rangle}  Z_Q[B]= (-1)^{\langle \nu_k, B\rangle} Z_Q[B]
\end{equation} where $\nu_k$ is the Wu class.
This apparent projectivity cannot be considered as coming from an invertible phase \emph{independent of} $B$.

We note that $PSL(2,\bZ)$ has a presentation \begin{equation}
PSL(2,\bZ)=\langle S,T\mid S^2=(ST)^3=1 \rangle  
\end{equation} and then \begin{equation}
PSL(2,\bZ_n)=\langle S,T\mid S^2=(ST)^3=1, \ T^n=1 \rangle.
\end{equation}
Therefore our relations \eqref{mainrelations} determine a projective action of $PSL(2,\bZ)$ 
and $PSL(2,\bZ_4)$.

Another closely related group is \begin{equation}
Mp(2,\bZ)=\langle S,T,C\mid S^2=C, (ST)^3=C^2, CT=TC, C^4=1\rangle
\end{equation} which is a double cover of \begin{equation}
SL(2,\bZ)=\langle S,T,C\mid S^2=C, (ST)^3=C^2, CT=TC, C^2=1\rangle.
\end{equation}
In many cases, the $SL(2,\bZ)$ action on theories with fermions should better be thought of as having an action of $Mp(2,\bZ)$, where $C^2$ is identified with $(-1)^F$, see \cite{Pantev:2016nze} and \cite[Sec.~8]{Hsieh:2020jpj}.
This is an extension of $SL(2,\bZ)$ by the fermion parity operator $(-1)^F$,
whereas our main relations \eqref{mainrelations} can be considered as an extension of $PSL(2,\bZ)$ by $Y$.
It seems  better to keep these two types of extensions conceptually separated,
since the fermion parity is part of the structure group of the spacetime
whereas $Y$ is an invertible phase in one dimension lower.

\subsection{Examples}

Let us mention a few examples of the dimension $d=2k$ and the structure $\str$ where the quadratic refinement is available.

\subsubsection{Oriented manifolds}
In arbitrary even dimensions $d=2k$, we can use the Pontrjagin square \begin{equation}
\mathcal{P}: H^k(M_{2k},\bZ_2)\to H^{2k}(M_{2k},\bZ_4),
\end{equation} which satisfies \begin{equation}
\mathcal{P}(B+B')=
\mathcal{P}(B)+
\mathcal{P}(B')+
2(B\cup B').
\end{equation}
Assuming that we equip the manifold with an orientation,
we can define the $\bZ_4$-valued quadratic refinement by \begin{equation}
q(B):= \int_{M_{2k}} \mathcal{P}(B).
\end{equation}

In this case, the corresponding Brown-Kervaire invariant is known \cite{Morita} to be the signature  modulo 8, \begin{equation}
Z_Y=\frac1{|H^k(M_{2k})|^{1/2}}\sum_b i^{\int_{M_{2k}}\mathcal{P}(b)}
= \exp\left(2\pi i \cdot \frac\sigma 8\right).
\end{equation}
This invertible phase is divisible by an arbitrary nonzero integer since $\sigma$ is $\bZ$-valued.
This allows us to remove the projectivity from the $SL(2,\bZ)$ action \begin{equation}
S^2=1,\quad (ST)^3=Y
\end{equation} by redefining $\tilde T:=TY^{-1/3}$.
This redefinition, however, is not guaranteed to keep the order of $T$ to be two or four.

\subsubsection{4d spin manifolds}
If we restrict attention to 4d spin manifolds,  the projectivity is trivial on the nose, since the signature is a multiple of eight.
Then the $SL(2,\bZ)$ operation is genuine and not projective.
Furthermore, since $S^2=T^2=1$, the operations in fact form the group $SL(2,\bZ_2)$.

This fact was used in \cite[Sec.~6]{Gaiotto:2014kfa} to refine the action of the Seiberg duality  on $\mathfrak{so}$ SQCD. 
Let us quickly review this here.
Recall first that in \cite{Intriligator:1995id} it was found that $\mathfrak{so}(N_c)$ theory with $N_f$ flavors is dual to $\mathfrak{so}(N_c')$ theory with the same number $N_f$ of flavors, where $N_c'=N_f-N_c+4$.
Later, in \cite{Aharony:2013hda}, it was noticed that there are in fact $Spin$, $SO_+$ and $SO_-$ theories, all sharing the same gauge algebra, and that the Seiberg duality acts as \begin{equation}
\begin{array}{ccc}
	Spin(N_c) & \leftrightarrow & SO_-(N_c'),\\
	SO_+(N_c) & \leftrightarrow & SO_+(N_c'),\\
	SO_-(N_c) & \leftrightarrow & Spin(N_c').
\end{array}
\end{equation}
This mapping however neglects the coupling of the theories to the background for the $\bZ_2$ 1-form symmetry.
Requiring that the $SL(2,\bZ)$ action to be compatible with the Seiberg duality, one finds that the mapping should in fact be \begin{equation}
\begin{array}{ccc}
	Spin(N_c) & \leftrightarrow & T(SO_-(N_c')),\\
	SO_+(N_c) & \leftrightarrow & T(SO_+(N_c')),\\ 
	SO_-(N_c) & \leftrightarrow & T(Spin(N_c')).
\end{array}
\end{equation} 
Equivalently, denoting $Q:=Spin(N_c)$ and $Q':=Spin(N_c')$ and introducing the symbol $\mathfrak{S}$ for the Seiberg duality, we can rewrite the mapping as \begin{equation}
\begin{array}{rcr}
	\mathfrak{S}Q & = & TSTQ',\\
	\mathfrak{S}SQ & = & TSQ',\\
	\mathfrak{S}STQ & = & TQ'.
\end{array}
\end{equation}
We can then check that $\mathfrak{S}$ commutes with the $SL(2,\bZ_2)$ action,
and that $\mathfrak{S}$ is of order 2, in the sense that \begin{equation}
\mathfrak{S}\mathfrak{S}Q
=\mathfrak{S}TSTQ'=TST\mathfrak{S}Q'
=(TST)^2Q=Q.
\end{equation}

\subsubsection{2d pin$^-$ manifolds}
In two dimensions, every manifold admits a pin$^-$ structure,
which is known to be in one-to-one correspondence with the $\bZ_4$-valued quadratic refinement $q$.
This can either be described using algebraic topology (see e.g.~\cite{KirbyTaylor}) or differential geometry (see e.g.~\cite[Appendix D]{Kaidi:2019tyf}).

The corresponding Brown-Kervaire invariant is often called \emph{the} Arf-Brown-Kervaire (ABK) invariant in recent physics literature,
and is known to generate the group $\Hom(\Omega_2^\text{pin$^-$},U(1))=\bZ_8$ of 2d pin$^-$ invertible phases.
It has been described as a fully-extended TQFT in \cite{DebrayGunningham}.
It also describes the low-energy limit of the time-reversal-invariant Kitaev Majorana chain, whose mod-8 behavior was first understood in \cite{Fidkowski:2009dba}.

The projectivity can be removed from the $PSL(2,\bZ)$ action by redefining $\tilde T=TY^5$.
It is impossible to do so from the $PSL(2,\bZ_4)$, since $\tilde S=SY^a$ and $\tilde T=TY^b$ would have to solve $2a=0$, $4b=0$ and $3(a+b)=1$ mod 8.

We can also restrict our attention to spin manifolds.
Then the theory $Y$ is the Arf theory and is $\bZ_2$-valued.
The projectivity can then be easily removed by redefining $\tilde T=TY$.

\subsubsection{4d pin$^+$ manifolds}
In four dimensions, (possibly non-orientable) manifolds with $w_2=0$ admit pin$^+$ structure,
which is known to be in one-to-one correspondence with the $\bZ_4$-valued quadratic refinement $q$.
This has a differential geometric description by realizing $B\in H^2(M_4)$ as the Stiefel-Whitney class $w_2$ of an $SO(3)$ bundle\footnote{%
Any element $B\in H^2(M_4)$ is the $w_2$ of an $SO(3)$ bundle.
To see this, we need to lift a map $B:M_4\to K(\bZ_2,2)$ to $B:M_4\to BSO(3)$ along $w_2: BSO(3)\to K(\bZ_2,2)$.
The obstruction is controlled by the fibration $BSU(2)\to BSO(3)\to K(\bZ_2,2)$.
Since $BSU(2)$ is 3-connected, the first obstruction to this lifting problem occurs when extending a map from the 4-skeleton to the 5-skeleton, 
and therefore no problem arises on 4-manifolds.
The authors thank Zheyan Wan and Juven Wang for the explanation.
\label{foot}
} and considering the associated eta invariant \cite[footnote 7,8]{Wan:2018bns}.

To determine the corresponding Brown-Kervaire invariant,
recall that $\Omega_4^\text{pin$^+$}=\bZ_{16}$ and that it is generated by $\mathbb{RP}^4$ \cite{KirbyTaylor}.
The generator $X$ of the group of the 4d pin$^+$ invertible phase, the massive Majorana fermion theory,
takes the values $Z_X=\exp(\pm 2\pi i \cdot \frac1{16})$.
This can be seen from the computation of the eta invariant \cite{Stolz}
and also from physical considerations \cite{Fidkowski:2013jua,Wang:2014lca,Metlitski:2014xqa,KitaevCollapse,Morimoto:2015lua}.

There are two pin$^+$ structures on $\mathbb{RP}^4$ and they correspond to two quadratic refinements of the intersection pairing on $H^2(\mathbb{RP}^4)=\bZ_2$.
We easily see that $Z_Y=\exp(\pm 2\pi i \cdot \frac18)$ there,
meaning that our $Y$ is twice the generator, $Y=X^2$.
As in the 2d pin$^-$ case, it is possible to remove the projectivity from the $PSL(2,\bZ)$ action but impossible to do so from the $PSL(2,\bZ_4)$ action.

\subsubsection{7d spin manifolds}
In passing, we mention that Witten's $SL(2,\bZ)$ action on 3d spin QFTs with $U(1)$ symmetry 
can be extended almost verbatim to an $SL(2,\bZ)$ action on 7d spin QFTs with $U(1)$ 2-form symmetry.
The $S$ and $T$ operations are given as in the 3d $U(1)$ case \eqref{U1S} and \eqref{U1T}:
\begin{align}
Z_{SQ}[C] &= \int [Dc] \exp\left(2\pi i  \int_{M_7} \frac{c}{2\pi} d \frac{C}{2\pi}\right) Z_Q[c],\\
Z_{TQ}[C] &= \exp\left(2\pi i \cdot \frac12 \int_{M_7} \left(\frac{C}{2\pi} d \frac{C}{2\pi} - \frac{p_1}2 d \frac{C}{2\pi} \right) \right) Z_Q[C].
\end{align}
Here, $C$ and $c$ are now background and dynamical 3-form gauge fields respectively and
the quadratic refinement to be used in the $T$ operation was constructed originally in \cite[Sec.~3]{Witten:1996hc} and was recently revisited in \cite[Sec.~4.4]{Hsieh:2020jpj}.
We note that it is crucial to have the term $(p_1/4) d (C/2\pi)$ for the spin structure to determine the quadratic refinement.
The computation of $S^2$ and $(ST)^3$ formally goes in the same way as in the 3d case detailed in \cite{Witten:2003ya}, and we obtain \begin{equation}
S^2=-1,\qquad (ST)^3=Y
\end{equation}
where $-1$ denotes charge conjugation and $Y$ is the 7d invertible phase whose action is the quadratic refinement itself. 
This theory $Y$ is the anomaly theory for the 6d self-dual tensor theory,
and is 28 times a generator of the group of the 7d spin invertible phases.\footnote{%
To see this, we first note that $\Omega^\text{spin}_8=\bZ\oplus \bZ$, 
generated by $\mathbb{HP}^2$ and $L_8$, where $4L_8$ is spin bordant to $K3\times K3$, see e.g.~\cite{Milnor}.
They have Pontrjagin numbers 
$p_1^2(\mathbb{HP}^2)=4$,
$p_2(\mathbb{HP}^2)=7$
and 
$p_1^2(L_8)=1152$, 
$p_2(L_8)=576$,
respectively.
The anomaly polynomial $\alpha_\text{gravitino}$, $\alpha_\text{tensor, naive}$ and $\alpha_\text{fermion}$ 
of the 6d gravitino, self-dual tensor and spin-$\frac12$ fermion, coming from the one-loop analysis, can be found in many places, and are given by 
$(275p_1^2-980p_2)/5760$, 
$(16p_1^2-112p_2)/5760$, 
$(7p_1^2-4 p_2)/5760$, respectively.
The anomaly polynomial of the self-dual tensor is then given by $\alpha_\text{tensor}=\alpha_\text{tensor, naive}+(p_1/4)^2/2=28\alpha_\text{fermion}$.
A short computation gives $\alpha_\text{fermion}(\mathbb{HP}^2)=0$,
$\alpha_\text{fermion}(L_8)=1$;
$\alpha_\text{gravitino}(\mathbb{HP}^2)=-1$,
$\alpha_\text{gravitino}(L_8)=-43$. 
This means that the anomaly of the 6d  gravitino and that of the 6d fermion generate the group $\bZ\oplus \bZ$ of the 7d spin invertible phases,
and the anomaly of a self-dual tensor is 28 times that of a fermion.
}
Therefore the projectivity cannot be removed by a redefinition.

\section{Interpretation via the bulk gauge theory}
\label{sec:interpretation}
\subsection{Generalities}
\label{sec:gen}
\begin{figure}[h]
\centering
	\begin{tikzpicture}[auto, scale=1][baseline=(current bounding box.center)]
		\draw[red,fill,opacity=.5] (0,0) -- (0,3) -- (1,4) -- (1,1) -- (0,0);
		\draw (0,0) -- (-5,0);
		\draw (0,3) -- (-5,3);
		\draw (1,4) -- (-4,4);
		\draw[dashed, lightgray] (1,1) -- (-4,1);
		\node at (-3,1.5) {bulk $d$-dim. $G$ gauge theory};
		\node[red, align=center] (boundary) at (3,2) {$(d-1)$-dim. theory\\with $G$ symmetry};
	\end{tikzpicture}
\caption{The theory with $G$ symmetry lives on the boundary of the $G$ gauge theory
\label{fig:1}}
\end{figure}
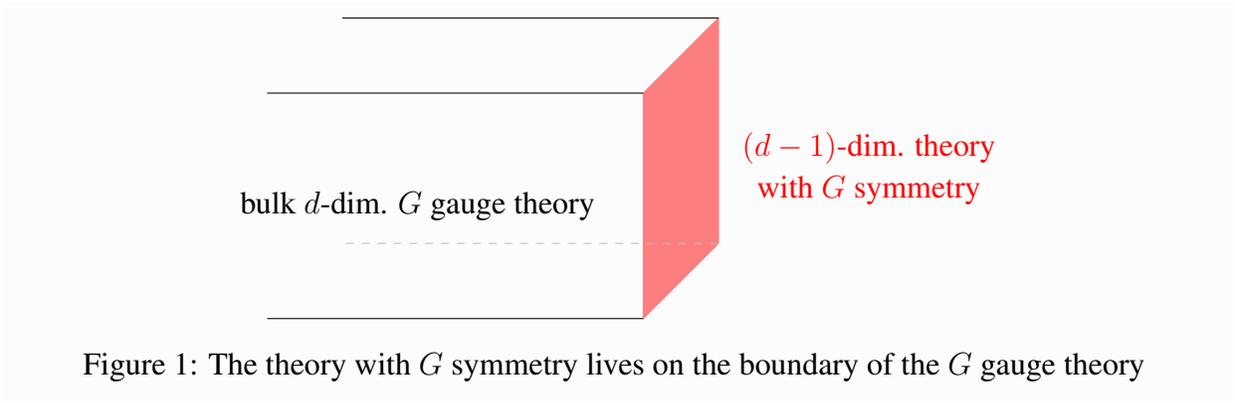

Let us now interpret our $SL(2,\bZ)$ action and its projectivity from the point of view of the bulk gauge theory.
We start with a general discussion, which is applicable to both Witten's original case and our cases.
Consider a $(d-1)$-dimensional theory with $G$ symmetry.
We attach to it a $d$-dimensional $G$ gauge theory with Neumann boundary condition for the gauge fields at the boundary, so that the $G$ symmetry on the boundary is gauged in this setup. See Fig.~\ref{fig:1}.

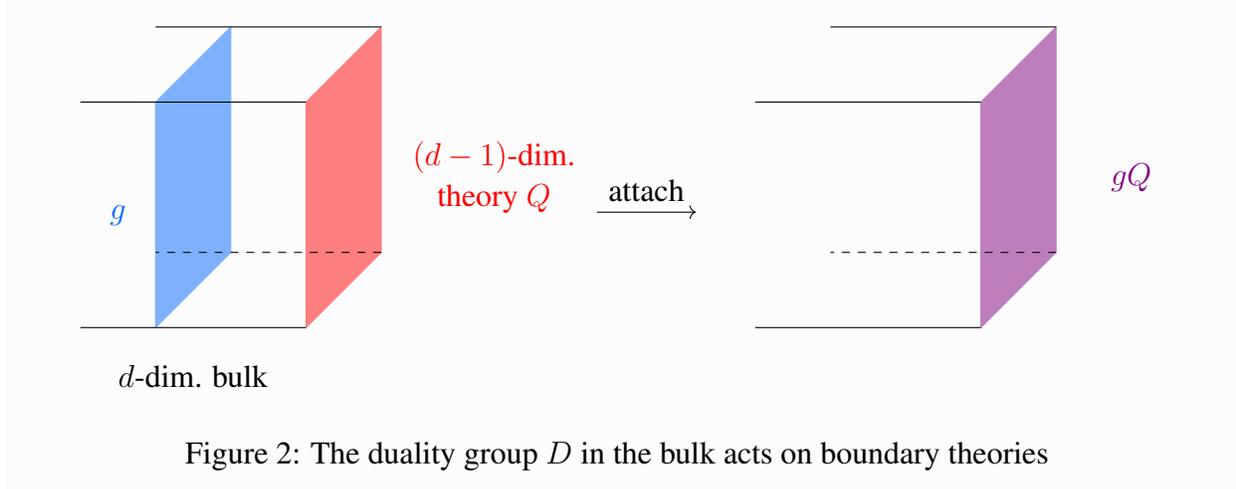
\begin{figure}[h]
\centering
	\begin{tikzpicture}
		\node (before) at (0,0) {
		\begin{tikzpicture}[auto, scale=1][baseline=(current bounding box.center)]
			\draw[red,fill,opacity=.5] (0,0) -- (0,3) -- (1,4) -- (1,1) -- (0,0);
			\draw[blue,fill,opacity=.5] (-2,0) -- (-2,3) -- (-1,4) -- (-1,1) -- (-2,0);
			\draw (0,0) -- node[below=10pt] {$d$-dim. bulk} (-3,0);
			\draw (0,3) -- (-3,3);
			\draw (1,4) -- (-2,4);
			\draw[dashed] (1,1) -- (-2,1);
			\node[red, align=center] (boundary) at (2.5,2) {$(d-1)$-dim.\\theory $Q$};
			\node[blue, align=center] (boundary) at (-2.5,1.5) {$g$};
		\end{tikzpicture}
		};
		\node (after) at (8,0) {
		\begin{tikzpicture}[auto, scale=1][baseline=(current bounding box.center)]
			\draw[violet,fill,opacity=.5] (0,0) -- (0,3) -- (1,4) -- (1,1) -- (0,0);
			\draw (0,0) -- node[below=10pt] {\phantom{$d$-dim. bulk}} (-3,0);
			\draw (0,3) -- (-3,3);
			\draw (1,4) -- (-2,4);
			\draw[dashed] (1,1) -- (-2,1);
			\node[violet, align=center] (boundary) at (2,2) {$gQ$};
			\node at (-3.5,0) {};
		\end{tikzpicture}
		};
		\draw[->] (before) -- node[above] {attach} (after);
	\end{tikzpicture}
\caption{The duality group $D$ in the bulk acts on boundary theories
\label{fig:2}}
\end{figure}

Let us suppose that the bulk $G$ gauge theory has a duality group $D$.
The duality group can depend on exactly which spacetime structure $\str$ we use;
we fix one particular $\str$, but do not make it explicit unless necessary.
The action of $D$ in the bulk can be represented by domain walls labeled by $g\in D$,
across which the duality operation $g$ is performed. 
We can now move this duality wall toward the boundary. 
This defines an action of the duality group $D$  on $(d-1)$-dimensional QFTs with $G$ symmetry.
See Fig.~\ref{fig:2}.

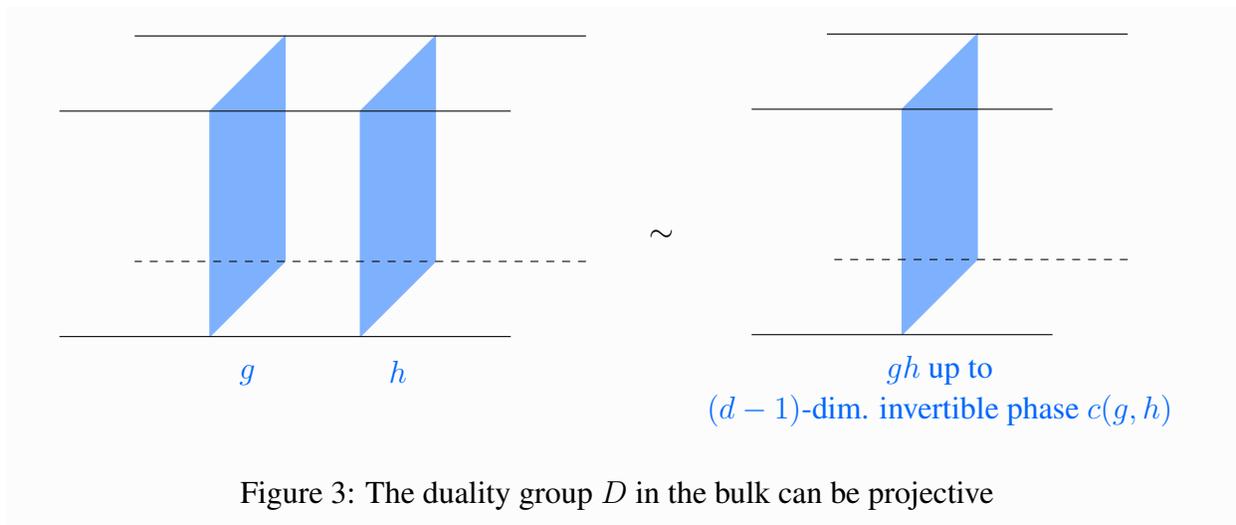
\begin{figure}[h]
\centering
	\begin{tikzpicture}
		\node (before) at (0,0) {
		\begin{tikzpicture}[auto, scale=1][baseline=(current bounding box.center)]
			\draw[blue,fill,opacity=.5] (0,0) -- (0,3) -- (1,4) -- (1,1) -- (0,0);
			\draw[blue,fill,opacity=.5] (-2,0) -- (-2,3) -- (-1,4) -- (-1,1) -- (-2,0);
			\draw (2,0) -- (-4,0);
			\draw (2,3) -- (-4,3);
			\draw (3,4) -- (-3,4);
			\draw[dashed] (3,1) -- (-3,1);
			\node[blue, align=center] at (0.5,-0.75) {$h$\\\phantom{a}};
			\node[blue, align=center] at (-1.5,-0.75) {$g$\\\phantom{a}};
		\end{tikzpicture}
		};
		\node (after) at (8,0) {
		\begin{tikzpicture}[auto, scale=1][baseline=(current bounding box.center)]
			\draw[blue,fill,opacity=.5] (0,0) -- (0,3) -- (1,4) -- (1,1) -- (0,0);
			\draw (2,0) -- node[below=10pt] {\phantom{$d$-dim. bulk}} (-2,0);
			\draw (2,3) -- (-2,3);
			\draw (3,4) -- (-1,4);
			\draw[dashed] (3,1) -- (-1,1);
			\node[blue, align=center] at (0.5,-0.75) {$gh$ up to\\ $(d-1)$-dim. invertible phase $c(g,h)$};
			\node at (-3,0) {};
		\end{tikzpicture}
		};
		\node at (4.5,0) {$\sim$};
	\end{tikzpicture}
\caption{The duality group $D$ in the bulk can be projective
\label{fig:3}}
\end{figure}

Now, the duality group action in the bulk can be anomalous in many ways.
One type of anomaly manifests itself as a failure of the composition law of the domain walls implementing the action of $D$, where it holds
only up to a multiplication by a $(d-1)$-dimensional invertible phase,
see Fig.~\ref{fig:3}.

Denoting the group of invertible phases  in dimension $d$  with structure $\str$ by $\Inv_\str^d$, such projectivity is
characterized by a 2-cocycle $c(g,h)\in \Inv_\str^{d-1}$,
and its cohomology class takes values in $H^2(BD,\Inv_\str^{d-1})$.
This projectivity in the bulk, when pushed on to the boundary,
then becomes the projectivity of the $D$ action on $(d-1)$-dimensional QFTs with $G$ symmetry.

The associativity of the composition of the domain walls can also fail, which can be parametrized by $H^3(D,\Inv_\str^{d-2})$. 
There can also be failures of higher associativities, given by $H^p(D,\Inv_\str^q)$ with $p+q=d+1$.
All these data comprise the $E_2$ page of the Atiyah-Hirzebruch spectral sequence converging to the group of invertible phases encoding the $D$ anomaly \cite{Gaiotto:2017zba,Thorngren:2018bhj}, \begin{equation}
E_2^{p,q}=H^p(BD,\Inv_\str^q) \Rightarrow \Inv_\str^{p+q}(BD).
\label{E2}
\end{equation}
In this paper, we only consider the effect of the part $(p,q)=(2,d-1)$.

Here, it seems appropriate to revisit the issue discussed in Sec.~\ref{sec:comments},
namely that in many theories with fermions the duality group $D=SL(2,\bZ)$ is extended by the fermion parity operation $(-1)^F$ to be $D'=Mp(2,\bZ)$.
More precisely, the structure group of the spacetime has the form \begin{equation}
[\str\times D']/\{1,(-1)^F\}.\label{twisted-structure}
\end{equation} 
The Atiyah-Hirzebruch spectral sequence computing the anomaly is known to have the same $E_2$ page as given in \eqref{E2}, 
whereas the differentials are known to be affected by the twist of the structure group given in \eqref{twisted-structure}, see e.g.~\cite{Thorngren:2018bhj} or \cite[Sec.~6.4]{Kaidi:2019tyf}.
As our discussion in this section only sees the $(p,q)=(2,d-1)$ part of the $E_2$ page,
we are not sensitive enough to see the extension of the duality group by the fermion parity.

\subsection{$U(1)$ symmetry in three dimensions}
In the case of 3d spin QFTs with $U(1)$ symmetry,
the bulk 4d theory is the Maxwell theory.
The part of the anomaly of the duality described by $H^2(SL(2,\bZ),\Inv^3_\text{spin})$ was determined in \cite{Seiberg:2018ntt} by reinterpreting the computation in \cite{Witten:1995gf}.
We recalled above that $\Inv^3_\text{spin}=\bZ$, whose generator we denoted by $X$.
The cohomology group is then given by $H^2(SL(2,\bZ),\bZ)=\bZ_{12}$.
The conclusion of  \cite{Seiberg:2018ntt}  is that the anomaly of the Maxwell theory corresponds to $8\in \bZ_{12}$.


Let us check that this matches with the projectivity of the $SL(2,\bZ)$ action \eqref{U1anom}.
A convenient representative of the extensions\footnote{%
In passing we mention that the extension for $n=1$ is isomorphic to a braid group $B_3=\langle A,B \mid ABA=BAB\rangle$. 
Its center is isomorphic to $\bZ$ and is generated by $\tilde C:=(AB)^3$.
One then takes $\tilde S:=ABA=BAB$, $\tilde T=A$ and $X=(AB)^6$.
} corresponding to $n\in \bZ_{12}=H^2(SL(2,\bZ),\bZ)$ are \begin{equation}
\tilde S^4=(\tilde S\tilde T)^3=X^n.
\end{equation}
We can confirm that $n$ is defined modulo 12 by redefining $\tilde S':=\tilde SX^3$ and $\tilde T':=\tilde TX$.
When $n=4m$, we can make a different redefinition $\hat S=\tilde SX^{-m}$, $\hat T=\tilde T$
which then satisfies \begin{equation}
\hat S^4=1,\qquad (\hat S\hat T)^3=X^m.
\end{equation}
Let us now recall that the projectivity of the $SL(2,\bZ)$ action on 3d theories was given by \eqref{U1anom}, which corresponds to $m=2$ since our theory $Y$, the level-1 $U(1)$ Chern-Simons theory, corresponds to $Y=X^2$.
We thus conclude that the anomaly is indeed given by $n=4m=8 \in \bZ_{12}$.

\subsection{$\bZ_2$ $(k-1)$-form symmetry in $2k$ dimensions}
Let us next discuss the case of $SL(2,\bZ)$ action on $2k$-dimensional QFTs with $\bZ_2$ $(k-1)$-form symmetry.
In this case, the analysis of the $SL(2,\bZ)$ action on the bulk $\bZ_2$ gauge theory with the required spacetime structure is not yet available in the literature.
Here, we will only make a few comments concerning how to understand the $SL(2,\bZ)$ action from the bulk point of view.

Let us consider the boundary theory on $M_{2k}$.
The bulk $\bZ_2$ gauge theory assigns its Hilbert space $\cH$ to it, which we are going to describe.
For this purpose, we note that the bulk gauge theory has operators supported on elements of $H_{k}(M_{2k})$.
For $k=1$, these are worldlines of anyons, and there are three types of them, often denoted by $E$, $M$, and $EM$, each corresponding to Wilson lines, 't Hooft lines, and dyonic lines of $\bZ_2$ gauge theory.

\begin{figure}[h]
\centering
	\begin{tikzpicture}
		\node (before) at (0,0) {
		\begin{tikzpicture}[auto, scale=1][baseline=(current bounding box.center)]
			\draw[green] (-1,0.25) .. controls (0,1.5) and (-1.5,2.5) .. (-0.5,3.5);
			\draw[preaction={draw=black!1, line width=8pt}, blue] (-2.5,1.5) .. controls (-1,1.25) and (-1,2.25) .. (0.5,2);
			\draw[red,fill,opacity=.5] (0,0) -- (0,3) -- (1,4) -- (1,1) -- (0,0);
			\draw (0,0) -- (-3,0);
			\draw (0,3) -- (-3,3);
			\draw (1,4) -- (-2,4);
			\draw[dashed] (1,1) -- (-2,1);
			\node[green, align=center] at (-1.0,-0.5) {$E(B)$};
			\node[blue, align=center] at (-3.5,1.5) {$M(B')$};
		\end{tikzpicture}
		};
		\node (before) at (9,0) {
		\begin{tikzpicture}[auto, scale=1][baseline=(current bounding box.center)]
			\draw[blue] (-2.5,1.5) .. controls (-1,1.25) and (-1,2.25) .. (0.5,2);
			\draw[preaction={draw=black!1, line width=8pt}, green] (-1,0.25) .. controls (0,1.5) and (-1.5,2.5) .. (-0.5,3.5);
			\draw[red,fill,opacity=.5] (0,0) -- (0,3) -- (1,4) -- (1,1) -- (0,0);
			\draw (0,0) -- (-3,0);
			\draw (0,3) -- (-3,3);
			\draw (1,4) -- (-2,4);
			\draw[dashed] (1,1) -- (-2,1);
			\node[green, align=center] at (-1.0,-0.5) {$E(B)$};
			\node[blue, align=center] at (-3.5,1.5) {$M(B')$};
		\end{tikzpicture}
		};
		\node at (5,0) {$= \quad (-1)^{\langle B, B'\rangle}$};
	\end{tikzpicture}
\caption{The braiding of bulk operators.
\label{fig:4}}
\end{figure}
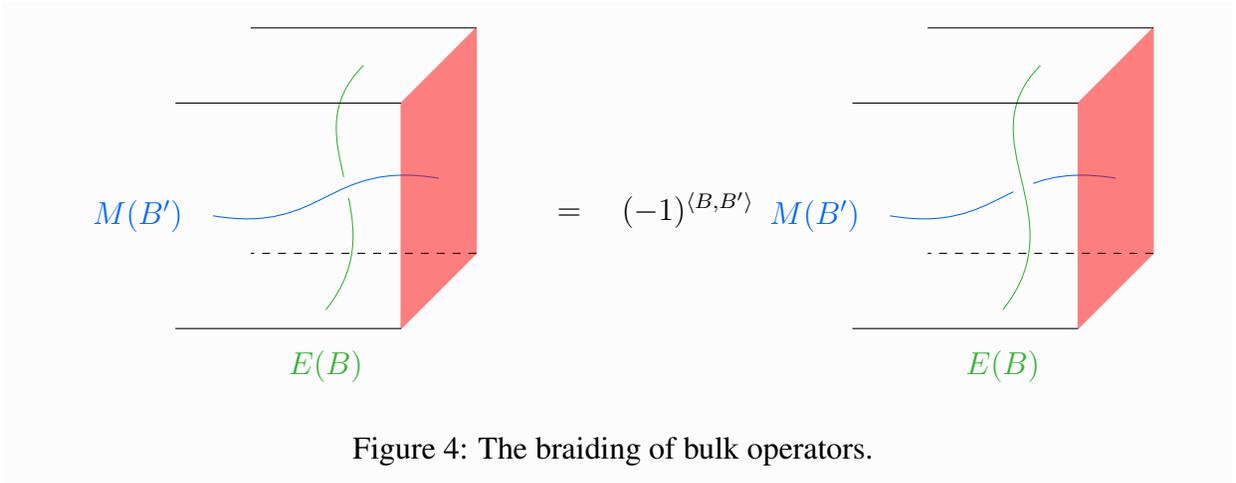

Generally, let us denote by $E(B)$ and $M(B)$ the operators of type $E$ and $M$ supported on the Poincar\'e dual of the element $B\in H^k(M_{2k})$.
We demand \begin{equation}
E(B)^2=M(B)^2=1.
\end{equation}
Furthermore, the braiding between them is given by \begin{equation}
E(B) M(B')= M(B') E(B) \cdot (-1)^{\langle B,B'\rangle},
\end{equation}
see Fig.~\ref{fig:4}.

The Hilbert space $\cH$ of the bulk theory forms an irreducible representation of this algebra.
We take the basis $\ket{B}$ where $E(B')$ is diagonalized by the relation \begin{equation}
E(B') \ket B = (-1)^{\langle B,B'\rangle} \ket B.
\end{equation} $M(B')$ then acts by the formula \begin{equation}
M(B') \ket B = \ket{B+B'}.
\end{equation}
We normalize the vectors $\ket B$ so that we have \begin{equation}
1=\frac{1}{|H^k(M_{2k})|^{1/2}}\sum_{B}\ket B \bra B.
\end{equation}

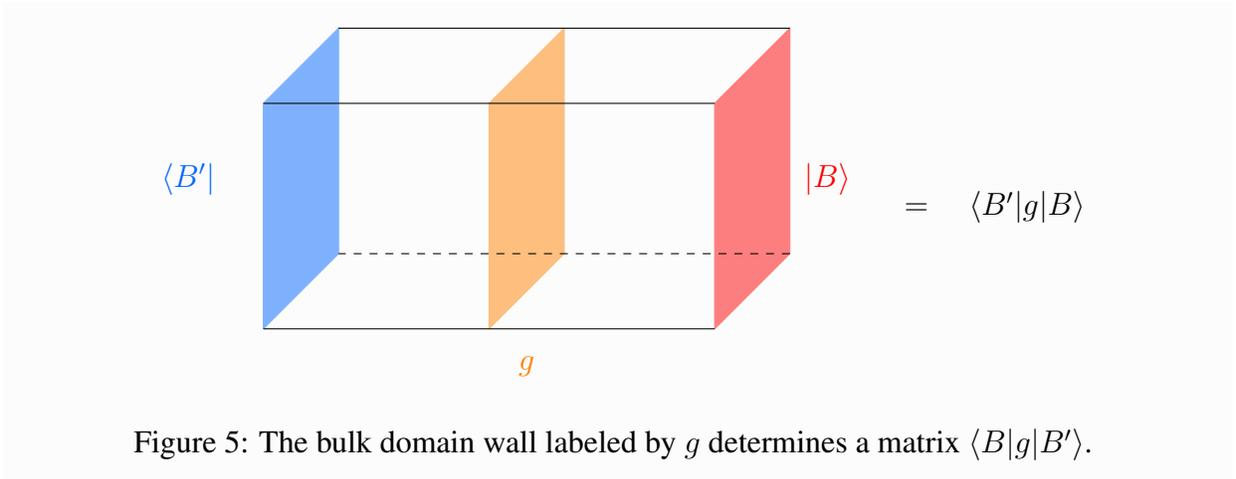
\begin{figure}[h]
\centering
	\begin{tikzpicture}
		\node (before) at (0,0) {
		\begin{tikzpicture}[auto, scale=1][baseline=(current bounding box.center)]
			\draw[red,fill,opacity=.5] (0,0) -- (0,3) -- (1,4) -- (1,1) -- (0,0);
			\draw[blue,fill,opacity=.5] (-6,0) -- (-6,3) -- (-5,4) -- (-5,1) -- (-6,0);
			\draw[orange, fill, opacity=0.5] (-3,0) -- (-3,3) -- (-2,4) -- (-2,1) -- (-3,0);
			\draw (0,0) -- (-6,0);
			\draw (0,3) -- (-6,3);
			\draw (1,4) -- (-5,4);
			\draw[dashed] (1,1) -- (-5,1);
			\node[red] at (1.5,2) {$\ket{B}$};
			\node[blue] at (-7,2) {$\bra{B'}$};
			\node[orange] at (-2.5,-0.5) {$g$};
		\end{tikzpicture}
		};
		\node at (6.5,0) {$= \quad \braket{B'|g|B}$};
	\end{tikzpicture}
\caption{The bulk domain wall labeled by $g$ determines a matrix $\braket{B|g|B'}$.
\label{fig:5}}
\end{figure}

We now regard the $2k$-dimensional theory $Q$ on the boundary to define a vector $\ket Q\in \cH$, so that \begin{equation}\label{GK}
Z_Q[B] = \braket{B|Q}.
\end{equation}
Similarly, the duality domain wall labeled by $g$ determines a matrix \begin{equation}
\braket{B'|g|B},
\end{equation}
see Fig.~\ref{fig:5}.

The $S$ operation in the bulk should exchange $E(B)$ and $M(B)$. 
This can be performed by a domain wall $S$ with the property \begin{equation}
\braket{B|S|B'}=(-1)^{\langle B,B'\rangle}.
\end{equation}
Indeed, it is easy to check \begin{equation}
SE(B)S^{-1}=M(B),\qquad SM(B)S^{-1}=E(B).
\end{equation}
This interpretation of the $S$ operation is long known to the experts, and can be found e.g.~in \cite[Lecture 4]{YTCernLect}.
It was also developed much more fully recently in \cite{Gaiotto:2020iye}.

Next we would like to discuss the $T$ operation.
Naively, we would like the $T$ operation in the bulk to keep $E(B)$ invariant and send $M(B)$ to $E(B)M(B)$.
There is a problem however, since \begin{equation}
(E(B)M(B))^2 = E(B)M(B)E(B)M(B) = (-1)^{\langle B,B\rangle}.
\end{equation}
To remedy this, we need to require the existence of the quadratic refinement, and consider the operator \begin{equation}
D(B):= i^{q(B)} E(B)M(B),
\end{equation} 
which satisfies \begin{equation}
D(B)^2=1
\end{equation} thanks to the relation \eqref{qBBB}.

We can regard the extra factor $i^{q(B)}$ as the expectation value of the transparent fermion line operator (and its higher-dimensional analogues) supported on the Poincar\'e dual of $B$.
In the case of $k=1$, this is analogous to the fact that $(-1)^{\bar q(B)}$ can be regarded as the expectation value of the transparent line operator for spin theories. 
The fact that this expectation value is $\bZ_4$-valued rather than $\bZ_2$-valued in the 2d/3d pin$^-$ case can be understood to follow from the fact that the reflection operation on the transparent fermion in a pin$^-$ theory squares to $-1$.
Consider the 3d pin$^-$ gauge theory on $\bR\times M_2$ where $M_2$ is a non-orientable 2-manifold, and wrap the transparent fermion on a cycle $B$ in $M_2$ which intersects the Poincar\'e dual $\mathcal{L}$ of $w_1(M_2)$.
In this setup, as the fermion crosses $\mathcal{L}$, it is acted upon by a reflection transformation. Since the reflection squares to $-1$, inserting another transparent fermion along $B$ (but separated from the previous fermion line along $\bR$) should leave us, after fusing the two fermion lines, with $-1$ times the setup such that no transparent fermion line wraps $B$ (at any point along $\bR$).
More concretely, the operator at the junction of fermion (wrapping $B$) and $\mathcal{L}$ squares to $-1$.

The above explanation implies that $i^{q(B)}=\pm1$  or $\pm i$  depending on whether  $\int_{M_2} B\cup w_1=0$ or $1$.
 In other words, it must be true that $2q(B)=2\int_{M_2} B\cup w_1$.
 Indeed, this can be verified by using (\ref{qBBB}) and $\langle B,B\rangle=\int_{M_2} B\cup w_1$.

This shows that it is important to take into account the effect of the transparent fermion line operator in 3d pin$^-$ theory and its higher-dimensional generalizations,
and its presence makes the $T$ operation to be of order four.

The $T$ operation can be performed by a domain wall $T$ with the property \begin{equation}
\braket{B|T|B'}=i^{q(B)} \delta_{B,B'}.
\end{equation}
It is again routine to check that \begin{equation}
TE(B)T^{-1}=E(B),\qquad TM(B)T^{-1}=D(B).
\end{equation}	
Our analysis in Sec.~\ref{sec:discreteSL2Z} then directly translates to the projectivity of the composition of the duality domain walls $S$ and $T$, which we do not repeat here.

It would be interesting to determine the full $SL(2,\bZ)$ anomaly of the 3d pin$^-$ $\bZ_2$ gauge theory.
As recalled in Sec.~\ref{sec:gen}, the projectivity we determined only fixes the part of the anomaly which is in $H^2(BSL(2,\bZ),\Inv^2_\text{pin$^-$})$ in the Atiyah-Hirzebruch spectral sequence.

Currently in the literature, a general theory of anomalies of  bosonic 3d TQFTs is available for both oriented \cite{Barkeshli:2014cna} and unoriented \cite{Barkeshli:2016mew} cases.
It is also known how to translate 3d spin and pin$^+$ theories to bosonic theories with an additional $\bZ_2$ symmetry \cite{Bhardwaj:2016clt,Bhardwaj:2016dtk}.
There is a steady progress in describing 3d spin TQFTs directly in terms of a spin version of modular tensor category, see e.g.~\cite{Bruillard:2016yio}.
The method available in the literature, therefore, cannot be directly applied to our question yet,
but it seems we are not very far away.
There might also be a different, more physics-based method to determine the $SL(2,\bZ)$ anomaly.
Either way, it would be a fun project to do so.


\section*{Acknowledgements}

YT thanks Chang-Tse Hsieh, Nati Seiberg, and Kazuya Yonekura for inspiring discussions while working on \cite{Seiberg:2018ntt,Hsieh:2020jpj}, from which the basic idea of this paper originated. 
The authors also thank Zheyan Wan and Juven Wang for the explanations concerning footnote \ref{foot}.
They also thank Eric Sharpe for a helpful comment on the v1 of the paper concerning the extension from $SL(2,\bZ)$ to $Mp(2,\bZ)$.

L.B.~is supported by ERC grants 682608 and 787185 under the European Union's Horizon 2020 programme. L.B.~also acknowledges support from NSF grant PHY-1719924.
Y.L.~is partially supported by the Programs for Leading Graduate Schools, MEXT, Japan, via the Leading Graduate Course for Frontiers of Mathematical Sciences and Physics
and also by JSPS Research Fellowship for Young Scientists.
Y.T.~is partially supported  by JSPS KAKENHI Grant-in-Aid (Wakate-A), No.17H04837 
and JSPS KAKENHI Grant-in-Aid (Kiban-S), No.16H06335,
and also by WPI Initiative, MEXT, Japan at IPMU, the University of Tokyo.

\def\arxivfont{\rm}
\bibliographystyle{ytamsalpha}
\baselineskip=.95\baselineskip
\bibliography{ref}

\newcommand{\etalchar}[1]{$^{#1}$}
\providecommand{\bysame}{\leavevmode\hbox to3em{\hrulefill}\thinspace}
\providecommand{\MR}{\relax\ifhmode\unskip\space\fi MR }
\providecommand{\MRhref}[2]{%
  \href{http://www.ams.org/mathscinet-getitem?mr=#1}{#2}
}
\providecommand{\href}[2]{#2}
\providecommand{\doihref}[2]{\href{#1}{#2}}
\providecommand{\arxivfont}{\tt}
\begin{thebibliography}{KTTW14}

\bibitem[ARS19]{Ang:2019txy}
J.~Ang, K.~Roumpedakis, and S.~Seifnashri, \emph{{Line Operators of Gauge
  Theories on Non-Spin Manifolds}},
  \doihref{http://dx.doi.org/10.1007/JHEP04(2020)087}{JHEP \textbf{04} (2020)
  087}, \href{http://arxiv.org/abs/1911.00589}{{\arxivfont arXiv:1911.00589
  [hep-th]}}.

\bibitem[AST13]{Aharony:2013hda}
O.~Aharony, N.~Seiberg, and Y.~Tachikawa, \emph{{Reading Between the Lines of
  Four-Dimensional Gauge Theories}},
  \doihref{http://dx.doi.org/10.1007/JHEP08(2013)115}{JHEP \textbf{08} (2013)
  115},
\href{http://arxiv.org/abs/1305.0318}{{\arxivfont arXiv:1305.0318 [hep-th]}}.

\bibitem[BBC{\etalchar{+}}16]{Barkeshli:2016mew}
M.~Barkeshli, P.~Bonderson, M.~Cheng, C.-M. Jian, and K.~Walker,
  \emph{{Reflection and Time Reversal Symmetry Enriched Topological Phases of
  Matter: Path Integrals, Non-Orientable Manifolds, and Anomalies}},
  \doihref{http://dx.doi.org/10.1007/s00220-019-03475-8}{Commun. Math. Phys.
  \textbf{374} (2019) 1021--1124},
  \href{http://arxiv.org/abs/1612.07792}{{\arxivfont arXiv:1612.07792
  [cond-mat.str-el]}}.

\bibitem[BBCW14]{Barkeshli:2014cna}
M.~Barkeshli, P.~Bonderson, M.~Cheng, and Z.~Wang, \emph{{Symmetry
  Fractionalization, Defects, and Gauging of Topological Phases}},
  \doihref{http://dx.doi.org/10.1103/PhysRevB.100.115147}{Phys. Rev. B
  \textbf{100} (2019) 115147},
  \href{http://arxiv.org/abs/1410.4540}{{\arxivfont arXiv:1410.4540
  [cond-mat.str-el]}}.

\bibitem[BGH{\etalchar{+}}16]{Bruillard:2016yio}
P.~Bruillard, C.~Galindo, T.~Hagge, S.-H. Ng, J.~Y. Plavnik, E.~C. Rowell, and
  Z.~Wang, \emph{{Fermionic Modular Categories and the 16-Fold Way}},
  \doihref{http://dx.doi.org/10.1063/1.4982048}{J. Math. Phys. \textbf{58}
  (2017) 041704}, \href{http://arxiv.org/abs/1603.09294}{{\arxivfont
  arXiv:1603.09294 [math.QA]}}.

\bibitem[BGK16]{Bhardwaj:2016clt}
L.~Bhardwaj, D.~Gaiotto, and A.~Kapustin, \emph{{State sum constructions of
  spin-TFTs and string net constructions of fermionic phases of matter}},
  \doihref{http://dx.doi.org/10.1007/JHEP04(2017)096}{JHEP \textbf{04} (2017)
  096}, \href{http://arxiv.org/abs/1605.01640}{{\arxivfont arXiv:1605.01640
  [cond-mat.str-el]}}.

\bibitem[Bha16]{Bhardwaj:2016dtk}
L.~Bhardwaj, \emph{{Unoriented 3D TFTs}},
  \doihref{http://dx.doi.org/10.1007/JHEP05(2017)048}{JHEP \textbf{05} (2017)
  048}, \href{http://arxiv.org/abs/1611.02728}{{\arxivfont arXiv:1611.02728
  [hep-th]}}.

\bibitem[BM05]{Belov:2005ze}
D.~Belov and G.~W. Moore, \emph{{Classification of Abelian Spin Chern-Simons
  Theories}}, \href{http://arxiv.org/abs/hep-th/0505235}{{\arxivfont
  arXiv:hep-th/0505235}}.

\bibitem[Bro72]{Brown}
E.~H. Brown, Jr., \emph{Generalizations of the {K}ervaire invariant},
  \doihref{http://dx.doi.org/10.2307/1970804}{Ann. of Math. (2) \textbf{95}
  (1972) 368--383}.

\bibitem[DG18]{DebrayGunningham}
A.~Debray and S.~Gunningham,
  \doihref{http://dx.doi.org/10.1090/conm/718/14478}{\emph{The {A}rf-{B}rown
  {TQFT} of pin{$^-$} surfaces}}, Topology and quantum theory in interaction,
  Contemp. Math., vol. 718, Amer. Math. Soc., Providence, RI, 2018, pp.~49--87.
  \href{http://arxiv.org/abs/1803.11183}{{\arxivfont arXiv:1803.11183
  [math-ph]}}.

\bibitem[FCV13]{Fidkowski:2013jua}
L.~Fidkowski, X.~Chen, and A.~Vishwanath, \emph{{Non-Abelian Topological Order
  on the Surface of a 3D Topological Superconductor from an Exactly Solved
  Model}}, \doihref{http://dx.doi.org/10.1103/PhysRevX.3.041016}{Phys. Rev.
  \textbf{X3} (2013) 041016},
\href{http://arxiv.org/abs/1305.5851}{{\arxivfont arXiv:1305.5851
  [cond-mat.str-el]}}.

\bibitem[FK09]{Fidkowski:2009dba}
L.~Fidkowski and A.~Kitaev, \emph{{The Effects of Interactions on the
  Topological Classification of Free Fermion Systems}},
  \doihref{http://dx.doi.org/10.1103/PhysRevB.81.134509}{Phys. Rev.
  \textbf{B81} (2010) 134509},
\href{http://arxiv.org/abs/0904.2197}{{\arxivfont arXiv:0904.2197
  [cond-mat.str-el]}}.

\bibitem[FM04]{Freed:2004yc}
D.~S. Freed and G.~W. Moore, \emph{{Setting the Quantum Integrand of
  M-theory}}, \doihref{http://dx.doi.org/10.1007/s00220-005-1482-7}{Commun.
  Math. Phys. \textbf{263} (2006) 89--132},
\href{http://arxiv.org/abs/hep-th/0409135}{{\arxivfont arXiv:hep-th/0409135}}.

\bibitem[GJF17]{Gaiotto:2017zba}
D.~Gaiotto and T.~Johnson-Freyd, \emph{{Symmetry Protected Topological Phases
  and Generalized Cohomology}},
  \doihref{http://dx.doi.org/10.1007/JHEP05(2019)007}{JHEP \textbf{05} (2019)
  007}, \href{http://arxiv.org/abs/1712.07950}{{\arxivfont arXiv:1712.07950
  [hep-th]}}.

\bibitem[GK20]{Gaiotto:2020iye}
D.~Gaiotto and J.~Kulp, \emph{{Orbifold Groupoids}},
  \href{http://arxiv.org/abs/2008.05960}{{\arxivfont arXiv:2008.05960
  [hep-th]}}.

\bibitem[GKSW14]{Gaiotto:2014kfa}
D.~Gaiotto, A.~Kapustin, N.~Seiberg, and B.~Willett, \emph{{Generalized Global
  Symmetries}}, \doihref{http://dx.doi.org/10.1007/JHEP02(2015)172}{JHEP
  \textbf{02} (2015) 172},
\href{http://arxiv.org/abs/1412.5148}{{\arxivfont arXiv:1412.5148 [hep-th]}}.

\bibitem[HTY20]{Hsieh:2020jpj}
C.-T. Hsieh, Y.~Tachikawa, and K.~Yonekura, \emph{{Anomaly inflow and $p$-form
  gauge theories}}, \href{http://arxiv.org/abs/2003.11550}{{\arxivfont
  arXiv:2003.11550 [hep-th]}}.

\bibitem[IS95]{Intriligator:1995id}
K.~A. Intriligator and N.~Seiberg, \emph{{Duality, Monopoles, Dyons,
  Confinement and Oblique Confinement in Supersymmetric SO(\hbox{$N_c$}) Gauge
  Theories}},
  \doihref{http://dx.doi.org/10.1016/0550-3213(95)00159-P}{Nucl.Phys.
  \textbf{B444} (1995) 125--160},
\href{http://arxiv.org/abs/hep-th/9503179}{{\arxivfont arXiv:hep-th/9503179}}.

\bibitem[Kit15]{KitaevCollapse}
A.~Kitaev, \emph{{Homotopy-Theoretic Approach to SPT Phases in Action:
  $\mathbb{Z}_{16}$ Classification of Three-Dimensional Superconductors}}.
  \url{http://www.ipam.ucla.edu/abstract/?tid=12389&pcode=STQ2015}.

\bibitem[Kla95]{KlausThesis}
S.~Klaus, \emph{Brown-{K}ervaire invariants}, Shaker, 1995.
  \url{https://www.researchgate.net/publication/267325025}. PhD thesis at
  Johannes Gutenberg-Universit\"at in Mainz.

\bibitem[KPMT19]{Kaidi:2019tyf}
J.~Kaidi, J.~Parra-Mart{\'\i ne}z, and Y.~Tachikawa, \emph{{Topological
  Superconductors on Superstring Worldsheets}},
  \doihref{http://dx.doi.org/10.21468/SciPostPhys.9.1.010}{SciPost Phys.
  \textbf{9} (2020) 10}, \href{http://arxiv.org/abs/1911.11780}{{\arxivfont
  arXiv:1911.11780 [hep-th]}}. {with a mathematical appendix by A.~Debray}.

\bibitem[KT90]{KirbyTaylor}
R.~C. Kirby and L.~R. Taylor, \emph{Pin structures on low-dimensional
  manifolds}, Geometry of Low-Dimensional Manifolds, vol.~2, London
  Mathematical Society Lecture Note Series, vol. 151, 1990, pp.~177--242.

\bibitem[KTTW14]{Kapustin:2014dxa}
A.~Kapustin, R.~Thorngren, A.~Turzillo, and Z.~Wang, \emph{{Fermionic Symmetry
  Protected Topological Phases and Cobordisms}},
  \doihref{http://dx.doi.org/10.1007/JHEP12(2015)052}{JHEP \textbf{12} (2015)
  052},
\href{http://arxiv.org/abs/1406.7329}{{\arxivfont arXiv:1406.7329
  [cond-mat.str-el]}}.

\bibitem[MFCV14]{Metlitski:2014xqa}
M.~A. Metlitski, L.~Fidkowski, X.~Chen, and A.~Vishwanath, \emph{{Interaction
  Effects on 3D Topological Superconductors: Surface Topological Order from
  Vortex Condensation, the 16 Fold Way and Fermionic Kramers Doublets}},
  \href{http://arxiv.org/abs/1406.3032}{{\arxivfont arXiv:1406.3032
  [cond-mat.str-el]}}.

\bibitem[MFM15]{Morimoto:2015lua}
T.~Morimoto, A.~Furusaki, and C.~Mudry, \emph{{Breakdown of the topological
  classification $\mathbb{Z}$ for gapped phases of noninteracting fermions by
  quartic interactions}},
  \doihref{http://dx.doi.org/10.1103/PhysRevB.92.125104}{Phys. Rev.
  \textbf{B92} (2015) 125104},
\href{http://arxiv.org/abs/1505.06341}{{\arxivfont arXiv:1505.06341
  [cond-mat.str-el]}}.

\bibitem[Mil63]{Milnor}
J.~Milnor, \emph{Spin structures on manifolds},
  \href{https://www.e-periodica.ch/digbib/view?pid=ens-001:1963:9::292}{Enseign.
  Math. (2) \textbf{9} (1963) 198--203}.

\bibitem[Mor71]{Morita}
S.~Morita, \emph{On the {P}ontrjagin square and the signature},
  \doihref{http://dx.doi.org/10.15083/00039814}{J. Fac. Sci. Univ. Tokyo Sect.
  IA Math. \textbf{18} (1971) 405--414}.

\bibitem[PS16]{Pantev:2016nze}
T.~Pantev and E.~Sharpe, \emph{{Duality Group Actions on Fermions}},
  \doihref{http://dx.doi.org/10.1007/JHEP11(2016)171}{JHEP \textbf{11} (2016)
  171},
\href{http://arxiv.org/abs/1609.00011}{{\arxivfont arXiv:1609.00011 [hep-th]}}.

\bibitem[RG99]{Read:1999fn}
N.~Read and D.~Green, \emph{{Paired States of Fermions in Two-Dimensions with
  Breaking of Parity and Time Reversal Symmetries, and the Fractional Quantum
  Hall Effect}}, \doihref{http://dx.doi.org/10.1103/PhysRevB.61.10267}{Phys.
  Rev. B \textbf{61} (2000) 10267},
  \href{http://arxiv.org/abs/cond-mat/9906453}{{\arxivfont
  arXiv:cond-mat/9906453}}.

\bibitem[Sto88]{Stolz}
S.~Stolz, \emph{Exotic structures on {$4$}-manifolds detected by spectral
  invariants}, \doihref{http://dx.doi.org/10.1007/BF01394348}{Invent. Math.
  \textbf{94} (1988) 147--162}.

\bibitem[STY18]{Seiberg:2018ntt}
N.~Seiberg, Y.~Tachikawa, and K.~Yonekura, \emph{{Anomalies of Duality Groups
  and Extended Conformal Manifolds}},
  \doihref{http://dx.doi.org/10.1093/ptep/pty069}{PTEP \textbf{2018} (2018)
  073B04},
\href{http://arxiv.org/abs/1803.07366}{{\arxivfont arXiv:1803.07366 [hep-th]}}.

\bibitem[Tac18]{YTCernLect}
Y.~Tachikawa, \emph{Topological phases and relativistic {QFTs}}.
  \url{https://member.ipmu.jp/yuji.tachikawa/lectures/2018-cern-rikkyo/}. Notes
  of the lectures given in the CERN winter school, February 2018.

\bibitem[Tho18]{Thorngren:2018bhj}
R.~Thorngren, \emph{{Anomalies and Bosonization}},
  \href{http://arxiv.org/abs/1810.04414}{{\arxivfont arXiv:1810.04414
  [cond-mat.str-el]}}.

\bibitem[Wit95]{Witten:1995gf}
E.~Witten, \emph{{On S Duality in Abelian Gauge Theory}},
  \doihref{http://dx.doi.org/10.1007/BF01671570}{Selecta Math. \textbf{1}
  (1995) 383},
\href{http://arxiv.org/abs/hep-th/9505186}{{\arxivfont arXiv:hep-th/9505186}}.

\bibitem[Wit96]{Witten:1996hc}
\bysame, \emph{{Five-Brane Effective Action in M-theory}},
  \doihref{http://dx.doi.org/10.1016/S0393-0440(97)80160-X}{J. Geom. Phys.
  \textbf{22} (1997) 103--133},
\href{http://arxiv.org/abs/hep-th/9610234}{{\arxivfont arXiv:hep-th/9610234}}.

\bibitem[Wit03]{Witten:2003ya}
\bysame,
  \doihref{http://dx.doi.org/10.1142/9789812775344_0028}{\emph{{$SL(2,Z)$
  Action on Three-Dimensional Conformal Field Theories with Abelian
  Symmetry}}}, From Fields to Strings: Circumnavigating Theoretical Physics
  (M.~Shifman, A.~Vainshtein, and J.~Wheater, eds.), World Scientific, 2005,
  pp.~1173--1200. \href{http://arxiv.org/abs/hep-th/0307041}{{\arxivfont
  arXiv:hep-th/0307041}}.

\bibitem[WS14]{Wang:2014lca}
C.~Wang and T.~Senthil, \emph{{Interacting Fermionic Topological
  Insulators/Superconductors in Three Dimensions}},
  \doihref{http://dx.doi.org/10.1103/PhysRevB.89.195124,
  10.1103/PhysRevB.91.239902}{Phys. Rev. \textbf{B89} (2014) 195124},
  \href{http://arxiv.org/abs/1401.1142}{{\arxivfont arXiv:1401.1142
  [cond-mat.str-el]}}.
[Erratum: Phys. Rev.B91,no.23,239902(2015)].

\bibitem[WW18]{Wan:2018bns}
Z.~Wan and J.~Wang, \emph{{Higher Anomalies, Higher Symmetries, and Cobordisms
  I: Classification of Higher-Symmetry-Protected Topological States and Their
  Boundary Fermionic/Bosonic Anomalies via a Generalized Cobordism Theory}},
  \doihref{http://dx.doi.org/10.4310/AMSA.2019.v4.n2.a2}{Ann. Math. Sci. Appl.
  \textbf{4} (2019) 107--311},
  \href{http://arxiv.org/abs/1812.11967}{{\arxivfont arXiv:1812.11967
  [hep-th]}}.

\bibitem[Yon18]{Yonekura:2018ufj}
K.~Yonekura, \emph{{On the Cobordism Classification of Symmetry Protected
  Topological Phases}},
  \doihref{http://dx.doi.org/10.1007/s00220-019-03439-y}{Commun. Math. Phys.
  \textbf{368} (2019) 1121--1173},
\href{http://arxiv.org/abs/1803.10796}{{\arxivfont arXiv:1803.10796 [hep-th]}}.

\end{thebibliography}

\end{document}